\documentclass{acmprocarticlesparxiv}

\usepackage{graphicx}

\begin{document}

% --- Author Metadata here ---
\permission{} % Allows default copyright data (0-89791-88-6/97/05) to be over-ridden - IF NEED BE.
% --- End of Author Metadata ---

\title{A Highly-Efficient Memory-Compression Scheme for GPU-Accelerated Intrusion Detection Systems}%

\numberofauthors{5} %  in this sample file, there are a *total*
\author{
% 1st. author
\alignauthor
Xavier J. A. Bellekens\\
       \affaddr{Department of Electronic and Electrical Engineering}\\
       \affaddr{University of Strathclyde}\\
       \affaddr{Glasgow, G1 1XW, UK\\xavier.bellekens@strath.ac.uk}
\alignauthor
Christos Tachtatzis\\
       \affaddr{Department of Electronic and Electrical Engineering}\\
       \affaddr{University of Strathclyde}\\
       \affaddr{Glasgow, G1 1XW, UK\\
       christos.tachtatzis@strath.ac.uk}
\alignauthor
Robert C. Atkinson\\
       \affaddr{Department of Electronic and Electrical Engineering}\\
       \affaddr{University of Strathclyde}\\
       \affaddr{Glasgow, G1 1XW, UK\\
       robert.atkinson@strath.ac.uk}
\and  
\alignauthor     
Craig Renfrew\\
       \affaddr{Keysight Technologies UK}\\
       \affaddr{5 Lochside Avenue}\\
       \affaddr{Edinburgh, EH12 9DJ, GB\\
       craig\_renfrew@keysight.com}
\alignauthor
Tony Kirkham\\
       \affaddr{Keysight Technologies UK}\\
       \affaddr{5 Lochside Avenue}\\
       \affaddr{Edinburgh, EH12 9DJ, GB\\
       tony\_kirkham@keysight.com}
}
% There's nothing stopping you putting the seventh, eighth, etc.
% author on the opening page (as the 'third row') but we ask,
% for aesthetic reasons that you place these 'additional authors'
% in the \additional authors block, viz.

% Just remember to make sure that the TOTAL number of authors
% is the number that will appear on the first page PLUS the
% number that will appear in the \additionalauthors section.
\maketitle
\begin{abstract}
Pattern Matching is a computationally intensive task used in many research fields and real world applications. Due to the ever-growing volume of data to be processed, and increasing link speeds, the number of patterns to be matched has risen significantly. In this paper we explore the parallel capabilities of modern General Purpose Graphics Processing Units (GPGPU) applications for high speed pattern matching. A highly compressed failure-less Aho-Corasick algorithm is presented for Intrusion Detection Systems on off-the-shelf hardware. This approach maximises the bandwidth for data transfers between the host and the Graphics Processing Unit (GPU). Experiments are performed on multiple alphabet sizes, demonstrating the capabilities of the library to be used in different research fields, while sustaining an adequate throughput for intrusion detection systems or DNA sequencing. The work also explores the performance impact of adequate prefix matching for alphabet sizes and varying pattern numbers achieving speeds up to 8Gbps and low memory consumption for intrusion detection systems.\\
\end{abstract}

\category{D.4.6}{Security and protection }[Information flow controls]
\category{K.6.5}{Management of Computing and Information Systems}{Security and Protection}

\keywords{Security, GPU, CUDA, Pattern Matching, Intrusion Detection Systems} % NOT required for Proceedings
\\
\section{Introduction}
With the exponential growth of networks, data storage, and network intrusions, the requirement for pattern matching, virus detection, and data categorisation poses a colossal challenge for internet service providers (ISP), and information technology (IT) administrators. Network Intrusion Detection Systems (NIDS) are one of the central pillars of an effective network defence system. As such, Intrusion Detection Systems (IDS) perform intensive pattern matching at strategic network points, or act as Host Intrusion Detection Systems (HIDS), scanning the incoming traffic for a particular host such as servers or end-host clients. 
The ever increasing prevalence of malware poses a significant threat to privacy and security, and the constant expansion of the link speed requires gigabit network scanning devices. \newline \newline
In essence, the challenge of an intrusion detection system, or a malware scanner, is similar to the problem of efficient pattern matching. Pattern matching is the most common task that requires most of time of the IDS, and therefore requires an efficient algorithm both in throughput and memory requirements \cite{Pungila} \cite{Pungila:2012:HMA:2403775.2403806}. \newline \newline
Typically, an IDS compare a large set of known-malicious \textit{signatures} against the potentially malicious data. Different approaches are used to cope with the large data sets, such as the one based on specialised hardware, like field-programmable gate arrays (FPGA) and application-specific integrated circuits (ASIC) \cite{1364659} \cite{FPGA1}. These hardware solutions achieve good performance, however,they are complex to customise, and are usually tied to a definitive application, making it difficult to update the design. \newline
On the other hand, commodity GPUs have shown their effectiveness at accelerating pattern matching operations for virus detection systems\cite{Gravity} \cite{5437720}, and network intrusion detection system \cite{bellekens} \cite{5683320}, while using off-the-shelf hardware and being easily customisable to the tasks required\cite{4041182}\cite{6299287}. Pushed forward by the video game industry, modern GPUs are constantly maturing, allowing flexible hardware capabilities, and powerful computational capabilities required by intrusion detection systems. The massive number of threads offered by the parallel capabilities of GPUs, allows them to overcome the problems induced by their CPU counterparts and simultaneously demonstrate better performance. \newline \newline 
This paper addresses the problems of dealing with large \textit{signature} data sets by proposing a highly-efficient memory compression scheme, applied to a highly parallelised multi-pattern matching algorithm, minimising the storage memory requirements and maximising the data transfers between the host and the GPU.\newline \newline
The contributions of this work are fourth fold:
\begin{itemize}
\item An efficient multi-pattern matching algorithm library is implemented and evaluated on GPUs. The algorithm can be transposed on any other multi-threaded system. 
\item A highly efficient memory-compression scheme is presented, by using bitmapping and a reduction algorithm on the trie. 
\item An Experimentation and analysis of the algorithm over multiple alphabets is realised, allowing the library to be used in different research fields. 
\item An implementation and thorough analysis of prefix pattern matching over different alphabet sizes is realised demonstrating significant speed increase in pattern matching. 
\end{itemize}
The remainder of this paper is organised as follows Section~\ref{sec:background} presents the background on general purpose GPU programming, and introduces multi-pattern matching, while in Section~\ref{sec:design_and_implementation} the storage model, the trie compression and the prefix matching are presented as well as the general approach and implementation. Furthermore Section~\ref{sec:performance_evaluation} shows the experimental results obtained with different scenarios, alphabet sizes, and prefix matching, and a comparison of the CPU and GPU throughput is realised. The paper ends with the conclusions and future work in Section~\ref{sec:conclusion}. 
 
%section
\section{Background}
\label{sec:background}
In this section basic concepts of GPUs, multi-pattern matching and related works are reviewed. 
\subsection{GPU Programming Model}
In this work the Nvidia Tesla K20m graphic card was selected, offering a rich Software Development Kit (SDK) also known as Computer Unified Device Architecture (CUDA) \cite{Nvidia}. The CUDA programming model allows researchers to harvest the massively parallel capabilities of generic consumer hardware using a flexible abstraction model through the Nvidia SDK. To communicate with the GPU, the C programming language has been extended with new libraries and directives, exposing the hardware features to the researchers. The major difference between CPUs and GPUs is based on the hardware layout. CPUs uses a single thread based on cache hierarchy focusing on reducing the latency for serial tasks, whereas GPUs focuses on increasing the raw processing power by dedicating the majority of its die surface to arrays of Arithmetic Logic Units (ALUs). The GPU is composed of several Streaming Multiprocessors (SM) operating in a Single Instruction Multiple Thread (SIMT) fashion and are composed of multiple Streaming Processors (SP) or CUDA cores. The number of SM and CUDA cores are architecture dependent.
\newline\newline
The code issued by the CPU to run on the GPU is called a \textit{kernel}, and is executed in five steps: (I) The host allocates space on the GPU to transfer the data; (II) The data is transferred from the host memory to the GPU memory via the DMA controller; (III) The host executes the kernel, instructing the GPU to execute the GPU code; (IV) The GPU executes the code in a massively parallel fashion; (V) The results are transferred back from the GPU memory to the CPU memory via the DMA controller. Kernels are executed in parallel with a finite number of threads. Threads are organised in thread blocks and each streaming multiprocessor executes one or more thread blocks. Within a block, threads are organised in \textit{warps}. Warps are a group of 32 threads, and are executed in a round robin fashion. A thread scheduler regularly switches from one warp to another, allowing the multiprocessor to maximise the resources of the GPU \cite{Sanders:2010:CEI:1891996}. \newline \newline Each SP within an SM shares an instruction unit, dedicated to the management of the instruction flow of the threads. When threads follow a different execution path, the overall throughput of the the SM is reduced, due to the serialisation of the instruction path. The divergent threads will be executed first until a common instruction is found, and the remaining threads will then converge to the same execution path \cite{kirk2012programming}. \newline \newline
Streaming multiprocessors share an off-chip L2 cache and possess their own set of registers. Each thread in a block can also share information via the \textit{shared memory}, while all threads launched can access data using the off-chip DRAM also called \textit{global memory}. Global memory requires numerous clock cycles to be accessed and is therefore expensive to use. Texture memory is a part of global memory, but is read only during the execution of the kernel and only allows defined types to be stored, however, the memory is accessed via a specific hardware, and data pulled are cached, allowing important speed-up when data requires numerous accesses~\cite{cook2012cuda}. \newline \newline
The Tesla K20m used in this work consists of 192 CUDA cores distributed over 19 multiprocessors, in addition to each SM the L1 cache 16-48KB of memory, and a shared memory of 16 to 48KB, both the L1 cache and shared memory must add up to 68KB in total as they are physically the same on-chip memory. The card is also composed of 5GB DRAM memory and consists of a single Printed Circuit Board (PCB).

\subsection{Multi-Pattern Matching}
Pattern matching consists of searching for one or more fixed patterns~$P$ in a body of text~$T$. Multi-string matching is widely used in intrusion detection systems, and virus detection engines. \newline
The Aho-Corasick (AC) Deterministic Finite state Automaton (DFA) is the most popular multi-pattern matching algorithm \cite{Aho:1975:ESM:360825.360855}. The concept of the AC algorithm is to maintain failure pointers that are invoked in the event of a mismatch at the current state. Failure pointers at the first level of the trie are always pointing back to the root node, however for the next levels the failure nodes are computed based on the longest prefix of the pattern currently matched. If no such prefix exists, the failure pointers point back to the root node. Each success pointer is associated with a label, corresponding to a character of the string and each states contains a boolean indicator expressing the status of the previous matched node. \newline 
The matching process starts at the root node and follows the transition states~$Q$, as long as the pattern searched for is present in the text~$T$. When no corresponding valid transition state is found the pointer invokes a failure transition, allowing to match every single pattern of the trie present in the text~$T$ in only one pass.
\newline \newline
In a multi-threaded environment the Aho-Corasick algorithm requires the text~$T$ to be split across multiple chunks, requiring each thread to analyse multiple chunks. This method has significant drawbacks, as a pattern may be span two chunks of data and the algorithm would not be able to match the pattern. Therefore overlapping is required, where each thread must scan the entire chunk of data and overlap the next chunk by the length of the longest pattern minus~1 \cite{5683320}\cite{bellekens}.
This method also reduces the number of threads allowed, as the number of threads launched is bound to the length of the input text and the maximum length of the pattern. This might create bottlenecks considering the massive number of threads able to be launched on a GPU. \newline
Another drawback of the AC algorithm in a multi-threaded environment, is the divergence created by the failure links, occurring when threads of the same warps execute different portions of the code, forcing the GPU to run several portions of the code in a serial fashion, and therefore reducing the overall throughput of the application~\cite{5683320}. 

\subsection{Parallel Failure-less Trie}
A substitute method to parallelise the Aho-Corasick algorithm has been designed and described by Lin~et~al.\cite{5683320}. The Parallel Failureless Aho-Corasick (PFAC) overcomes the problem of split patterns over two chunks of data, and reduces the divergence by removing the failure transitions pointers from the state machine and assigning each thread to a specific character of the text~$T$.\newline \newline
The parallel failure-less method harvests the full potential of the GPU by maximising the number of threads assigned to the text~$T$, and allows the first memory transfer of the text string to happen in a coalesced manner. Afterwards the threads will transition to separate execution paths leading to un-coalesced memory accesses and thread divergence, however, most of them will terminate at an early stage, as the assigned letter do not match the \textit{signatures}.
\newline
However, as the threads progresses deeper in the trie, more threads will be discarded due to mismatches. The remaining threads matching the patterns, leaving only a minority of threads matching, and therefore avoiding more thread divergence.

\subsection{Prefix Matching} 
To reduce the thread divergence occurring  after the first level of the trie, and speed up the matching process, the trie can be truncated to a depth of 5 for alphabets over 52 characters. This approach is known as prefix matching and assumes that a secondary process fully matches the patterns after the first matching stage \cite{938073}.\newline \newline 
This approach is effective due to the length of the malicious pattern and therefore has a low probability of being a false positive. Vasiliadis~et~al. \cite{Gravity} demonstrated that a prefix of length 8 is sufficient to produce a false positive rate below 0.0001\%. \newline \newline
Reducing the length of the pattern also reduces the thread divergence occurring at deeper levels of the trie, and reduces the memory footprint due to the number of states truncated.

\begin{figure}[t]
  \includegraphics[width=3.5in]{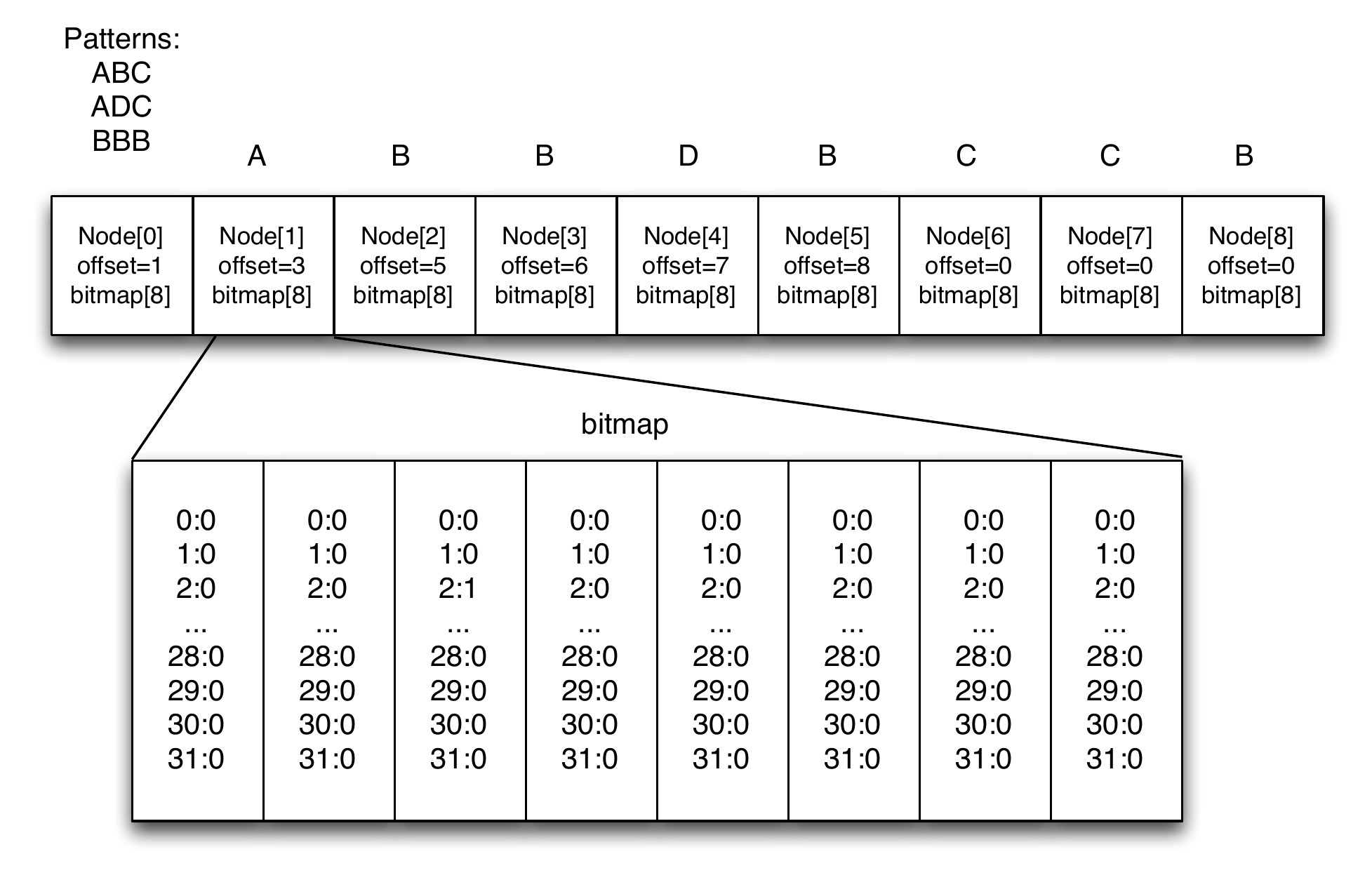}
  \caption{Breadth-first constructed trie, and the bitmap corresponding to the first node of the trie.}
\label{fig:bitmap}
\end{figure}
%section
\section{Design and Implementation}
\label{sec:design_and_implementation}
The library presented in this paper significantly reduces the memory footprint of the trie, while achieving high performances throughput. To accomplish this performance a failure-less Aho-Corasick algorithm has been implemented, along with a reduced data structure, a highly-efficient storage model, a trie compression algorithm and a prefix matching algorithm.
\subsection{Storage Model}
The construction of the trie is stack based, and relies on an array of structures, allowing the trie to be stored in a continuous memory block. The construction of the trie occurs in a breadth-first (BF) fashion, and each pattern is added to the trie after being sorted alphabetically. The final structure of the array has been stripped of unnecessary data and is composed of an offset pointing to first child of the current node, or to 0 if final. The children of each node are represented in a 256 bitmap, each bit corresponding to one letter of the ASCII alphabet. A population count on the bitmap is made to obtain the position of the child required.
\newline\newline
Figure~\ref{fig:bitmap} shows how the nodes are organised in contiguous memory. The bitmap of the node at cell 1 represents an array of 8 cells of 32 bits, representing the 256 ASCII alphabet. The first node `A' has two children `B' and `D' (based on the two first patterns). The offset of \textit{node[1]} only points to its first child (\textit{node[3]}) corresponding to the letter `B', however, to access the second node `D' at position \textit{node[4]}, a population count is required. The population count, sums the number of bits set to 1 in the bitmap, before the ASCII value of the letter required. E.g. The letter `B' corresponds to the ASCII value 66, and letter `D' corresponds to 68, meaning that the bits at position 66 and 68 are set to 1.  When an access to the node corresponding to the letter `D' is required, a population count, look up the number of bits set to one before the corresponding ASCII value of the letter `D' (68) . In this case only one bit is set, corresponding to the letter `B' at position 66. The transition is then calculated as follows :  let $i$ be the value of the population count, and \textit{offset} be the value of the first child.
\begin{equation}
curNode=curNode \rightarrow offset+i
\end{equation}
As $i$ is equal to 1 and the offset of the first node `A' corresponds to 3, the second child can be found at position 3+1~=4, corresponding in Figure~\ref{fig:bitmap} to node \textit{node[4]}.
\newline\newline
This approach, shows that the trie only requires  $4 +  32 = 36$~bytes per nodes, as the structure only needs 32~bytes for the bitmap and 4~bytes for the offset. \newline\newline
The bitmap can be adapted to the size of the alphabet and occupy~$|\Sigma|$ bits. Given an alphabet of $\Sigma=32$, and a corresponding bitmap size, it is possible to compare these results against PFAC, their structure requires 15 bytes per node for a total of 24.18~MB over 1,703,023~nodes in total. The approach described by Pungila et al. \cite{Pungila:2012:HMA:2403775.2403806}  using Aho-Corasick and the Commentz-Walter (AC-CW) algorithm requires 10~bytes per node for a total of 15.02~MB, while our approach only requires 32~bits of bitmap, and 32~bits of offset only allowing the trie to be stored in 12.9~MB requiring 1.16 times less memory than the AC-CW trie, and 1.87~times less memory than PFAC to store the complete trie in~memory. Comparing our method to GrAVity \cite{Gravity} with an alphabet $\Sigma=256$ and a bitmap of 256~bits our storage model for a total of 352,921 nodes only requires 12.1MB while GrAVity requires 345MB, which is 28.5 times more memory than our storage model. 

\begin{figure}[t!]
  \includegraphics[width=3.5in]{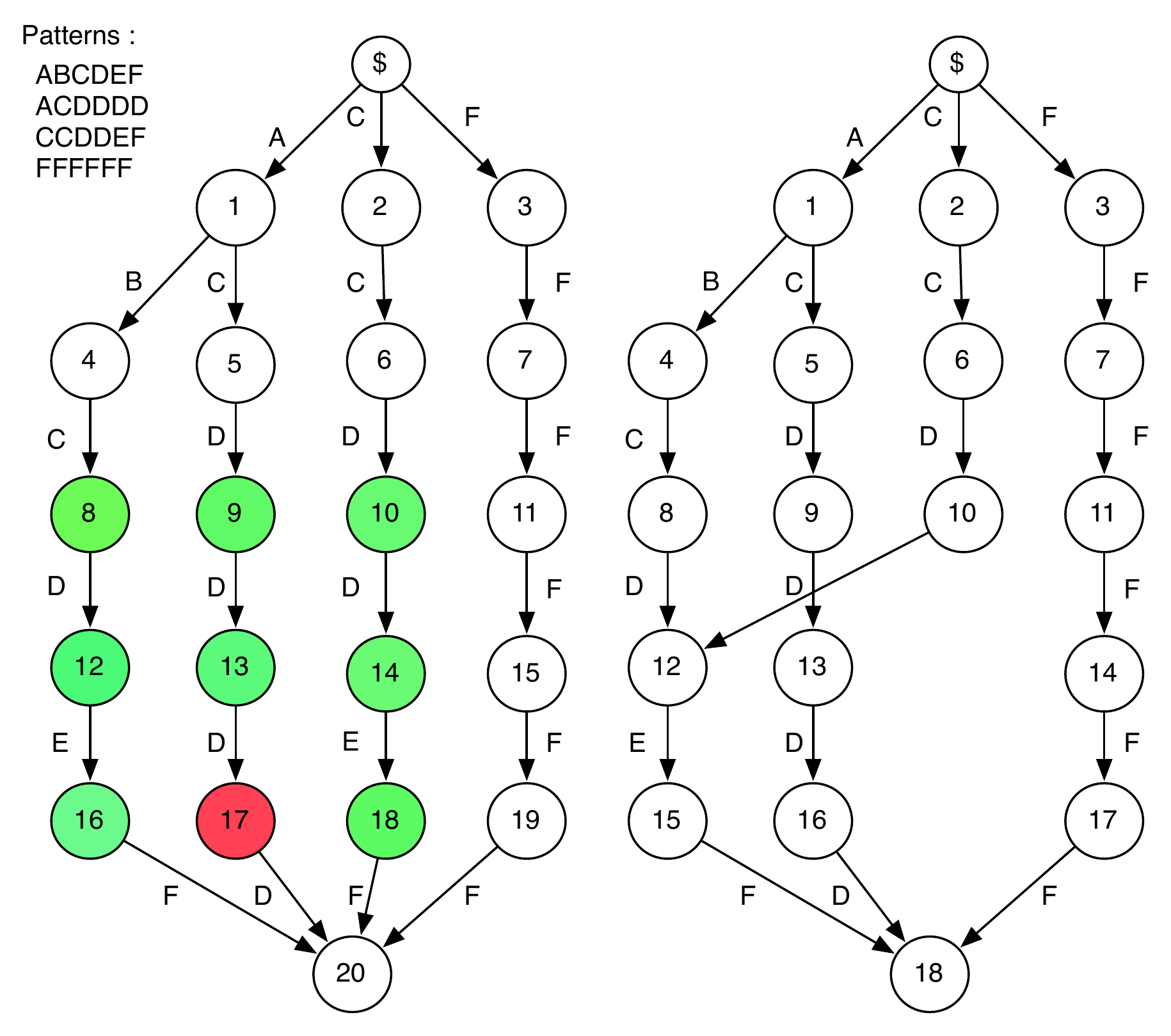}
  \caption{Trie node compression.}
\label{fig:nodeReduce}
\end{figure}
%This approach, shows that the trie only requires  $4 +  32 = 36$~bits per nodes, as the structure only needs 32~bits for the bitmap and 4~bits for the offset. Comparing this result to PFAC, the structures requires 15 bytes per node for a total of 24.18~MB over 1,703,023~nodes in total. The approach described by Pungila et al. \cite{Pungila:2012:HMA:2403775.2403806}  using Aho-Corasick and the Commentz-Walter (AC-CW) algorithm requires 47~bits per node for a total of 15.02~MB. Our approach only requires 7.3~MB, half the memory required by the AC-CW trie, and 3.3~times less memory than PFAC to store the complete trie in~memory. 

\subsection{Trie Compression}
To further minimise the size of the trie, a compression algorithm has been designed, allowing similar pattern suffixes to be merged, and improving the algorithm flexibility. The reduction process consists of two major steps. The first step consists of merging the last nodes of every pattern into a single one and can be calculated as follows, let $Q$ be the number of states and $P$ the number of patterns :
\begin{equation}
Q_{final}=Q  - (\mid P \mid -1)
\label{eq:eq_nodeReduce}
\end{equation}
Equation~\ref{eq:eq_nodeReduce} allows the final number of nodes after the first step of the reduction to be calculated. This reduction is possible as the final node of every pattern does not possess any children, and therefore only one global final node is required as shown in Figure~\ref{fig:nodeReduce}. \newline
The second step occurs on the last three levels of the trie, as these levels show less divergence in the suffixes of the patterns. Merging the last levels, allows the algorithm to significantly reduce the number of nodes. 

Figure~\ref{fig:nodeReduce} shows how the reduction occurs, where the last three levels of each pattern are compared against each other and merged when a match occurs. 
\newline \newline
The average reduction of the trie can also be calculated, considering that the trie contains $n$ patterns and of alphabet $\Sigma$, and that all letters are equiprobable and independent of each other. All the possible combination to construct a chain of 2 characters is therefore calculated as follows : 
\begin{equation}
r = \Sigma^2
\end{equation}
The average reduction can then be calculated as follows, let $r$ be the number of chains, $n$, the number of patterns $P$ the probability of a trie having exactly $i$ unique suffixes, and $L$ the average trie length after reduction : 
\begin{equation}
    \overline{L}=\sum\limits_{i=1}^{r} P(g=i)\cdot 2i = \sum\limits_{i=1}^{min(r,n)} \frac{\binom{r}{i}\left(\binom{n-i}{i}\right)}{\left(\binom{n}{r}\right)}\cdot 2i
\label{eq:eq_averageReduction}
\end{equation}
Equation~\ref{eq:eq_averageReduction} can be used to quantify the average reduction obtained by the reduction algorithm and evaluate the memory requirements of the trie on the GPU.
\begin{figure*}[!t]
  \includegraphics[width=6.9in]{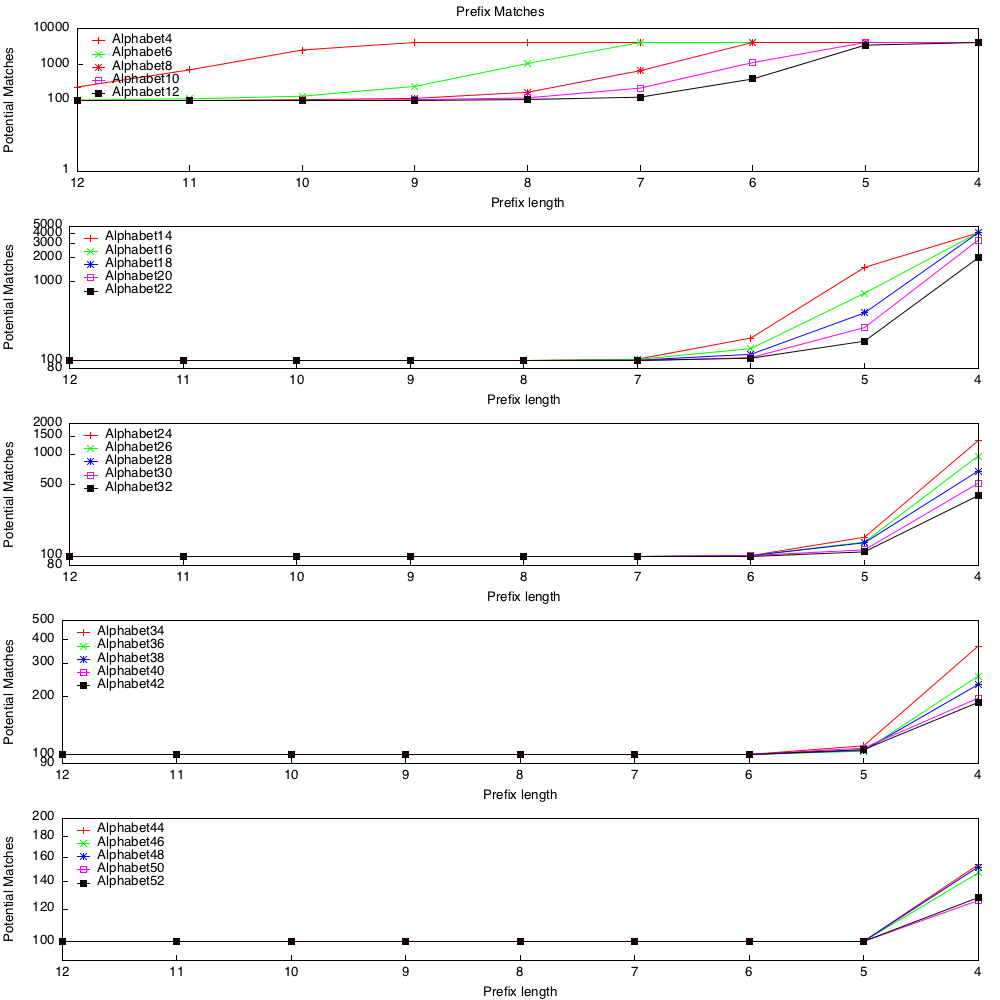}
  \caption{Prefix matching over different alphabets.}
\label{fig:prefixMatching}
\end{figure*}

\begin{figure}[!h]
  \includegraphics[width=3.5in]{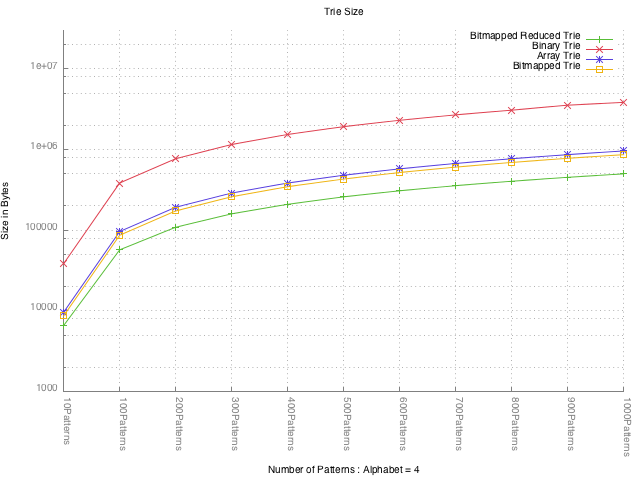}
  \caption{Trie size over a $\Sigma$ of 4.}
\label{fig:trieSize4}
\end{figure} 

\begin{figure}[b!]
  \includegraphics[width=3.5in]{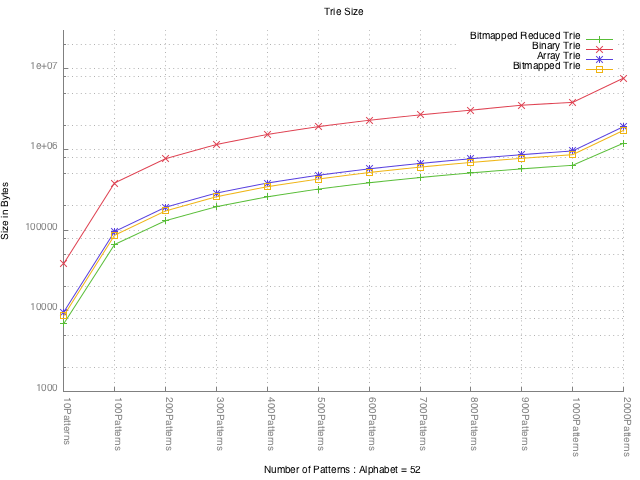}
  \caption{Trie size over a $\Sigma$ of 52.}
\label{fig:trieSize52}
\end{figure}

%section
\section{Performance Evaluation}
\label{sec:performance_evaluation}
The series of tests conducted to evaluate the performances of this approach have been realised using synthetic data sets randomly generated using Python and the Mersenne Twister (MT) uniform pseudo-random numbers generator \cite{Matsumoto:1998:MTE:272991.272995}.\newline\newline\newline
The evaluation of the algorithms is performed against 5 different data sets of exactly 100~MB each. The uniqueness of each file is ensured by computing the SHA256~hash of the files. Each algorithm is then run 100~times against each data set and the results presented are the average of the total run time. This technique permits to mitigate the sources of error, such as background processes requesting resources.\\\\
\subsection{Memory Hierarchies}
The trie and reduction algorithms are evaluated against different numbers of patterns and alphabet sizes. Figure~\ref{fig:trieSize4} demonstrates the performance improvements of the bitmapped reduced trie over the size of a simple bitmapped trie, a binary trie and an array trie with an alphabet $\Sigma=4$ and patterns of 20 characters. The reduction achieved by the bitmapped reduced trie demonstrates an impressive average size reduction improvement of 38\% compared to the bitmapped trie. 

Figure~\ref{fig:trieSize52} depicts the performance of the different types of trie against an alphabet $\Sigma=52$ and patterns of 20~chars. The increase of the alphabets size shows that the reduction algorithm is more effective when using smaller alphabets with synthetic files. This phenomenon is due to the lack of similar suffixes in the randomly generated patterns over larger alphabets.

\begin{figure}[t!]
  \includegraphics[width=3.5in]{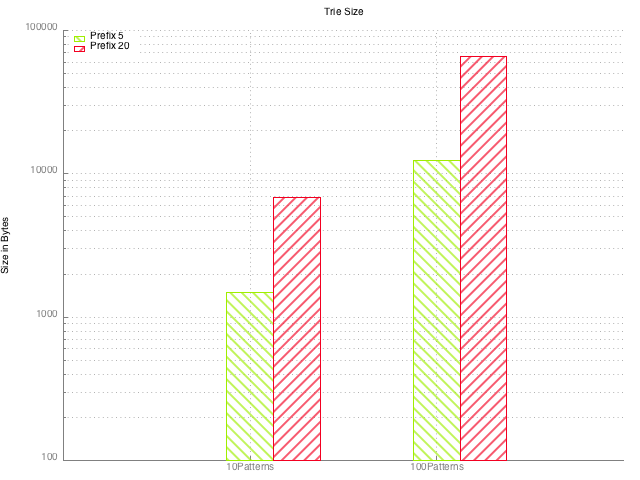}
  \caption{Prefix trie size comparison.}
\label{fig:prefixTrieSize}
\end{figure}
\subsection{Prefix Marching Evaluation}
The technique of prefix matching is alphabet dependent as the probability of a prefix being unique is smaller when smaller alphabets are used. Figure~\ref{fig:prefixMatching} demonstrates this statement. In this evaluation 10,000~threads are launched and the data set contains 100~patterns to be found. The patterns and the alphabets are composed of different alphabet sizes from $\Sigma=4$ to $\Sigma=52$.
\newline \newline  
Figure~\ref{fig:prefixMatching} shows that an alphabets of size $\Sigma=4$ will require an average prefix of over 12 characters, before being able to match the 100 patterns successfully. However, as the alphabet sizes rises up to $\Sigma=52$, the prefix length dramatically decreases. The prefix matching stabilises from $\Sigma=44$ up to larger alphabets and requires an average of 5 characters to be matched before the validation of the prefix matching can occur. These results verify the results of Vasiliadis et al.~\cite{Gravity} which required prefixes of 8 characters to successfully match the set of signature used in their study.
\newline \newline
Figure~\ref{fig:prefixTrieSize} shows the gain obtained by using a bitmapped reduced trie with prefix matching over an alphabet of $\Sigma=52$, for 10 and 100 patterns.

\subsection{Throughput Scaling Evaluation}
\begin{figure}[b!]
  \includegraphics[width=3.5in]{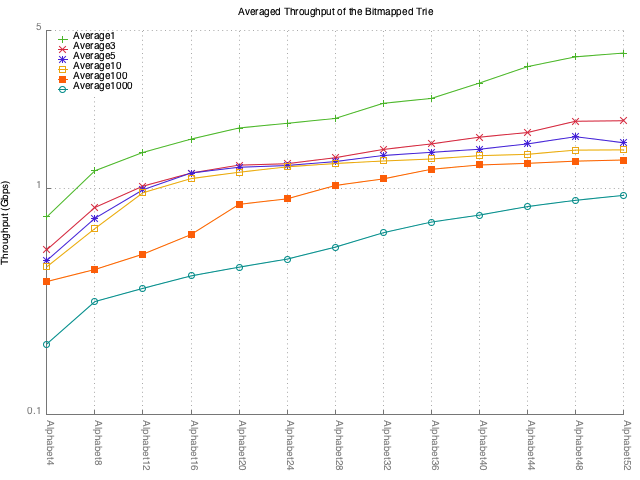}
  \caption{Throughput sustained for small alphabets.}
\label{fig:smalAlpha}
\end{figure}

In the next experiment the throughput of the algorithm is evaluated against different alphabet sizes, and number of patterns matched. Figure~\ref{fig:smalAlpha} shows the performance achieved by the algorithm against smaller alphabets, with a prefix matching according to the size of the alphabet $\Sigma$. As the alphabet size rises the throughput sustained by the matching process rises. This is due to the number of threads being cancelled at the first matched letter. Larger alphabets increases the number of cancelled threads, and reduce the divergence occurring during the matching process. When matching a thousand patterns, the throughput rises to 1~Gbps over an alphabet of 52, and when matching only one pattern, the throughput rises to 4.6~Gbps, demonstrating an impressive throughput for small alphabet sizes.

\begin{figure}[t!]
  \includegraphics[width=3.5in]{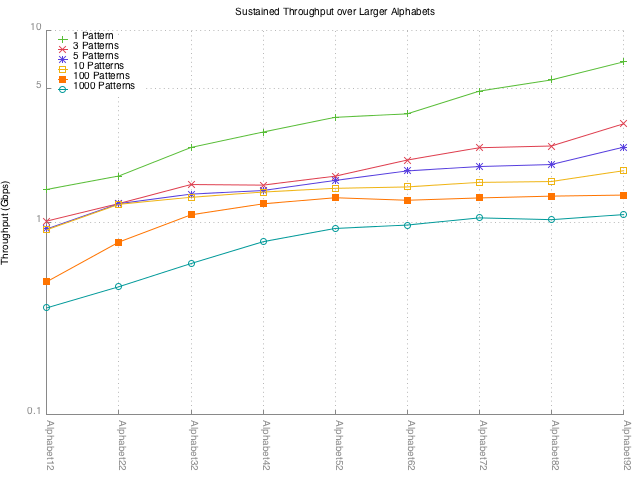}
  \caption{Throughput sustained for large alphabets.}
\label{fig:largeAlpha}
\end{figure}

Figure~\ref{fig:largeAlpha} demonstrates the behaviour of the algorithm over larger alphabet sizes over synthetic files. For one pattern matched the algorithm nearly reaches 8~Gbps and for a thousand patterns, the speeds stabilises after reaching an alphabet of $\Sigma=52$ to 1~Gbps. The number of patterns matched are influencing the throughput of the algorithm, as the divergence created by the higher number of patterns slows down the matching process, requiring the GPU to perform more sequential operations. 

\begin{figure}[b!]
  \includegraphics[width=3.5in]{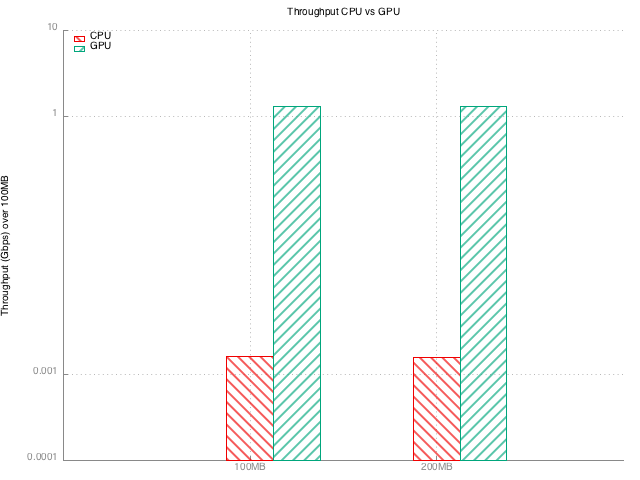}
  \caption{Comparison between the throughput of the CPU and the GPU.}
\label{fig:speedCPUGPU}
\end{figure}

Figure~\ref{fig:speedCPUGPU} shows a comparison of the throughput of CPUs and GPUs searching for 100 patterns, over an alphabet $\Sigma=52$ and files of 100~MB and 200~MB. The size of the file do not affect the overall throughput of the algorithm, and demonstrates a throughput of 1.3~Gbps for the GPU implementation of the bitmapped reduced failure less trie.
%section
\section{Conclusion}
\label{sec:conclusion}
In this work we have presented an efficient bitmapped reduced failure-less trie library for intrusion detection systems on GPUs and for applications requiring the use of smaller alphabets such as DNA sequencing. \newline \newline
The experimental results presented have shown that the memory-compression scheme in this work is ideal for GPU intrusion detection systems, using 1.87~times less memory than PFAC, over 1.16~times less memory than the AC-CW approach, and over 28.5~times less memory than what is required in the implementation of GrAVity. Our approach also achieved up to 8~Gbps and demonstrated a thorough analysis of prefix matching applied to off-the-shelf technology for varying alphabets. 

Future work include real time network packet scanner on off-the-shelf technology such as SNORT, taking advantage of multi-GPU hardware, and improvements to the library. 

\section*{Acknowledgment}
The authors would like to thank Keysight Technologies UK Ltd for their insightful comments and feedback as well as their support. 
\bibliographystyle{abbrv}
\bibliography{HEPFACArxiv}  % sigproc.bib is the name of the Bibliography in this case

\begin{thebibliography}{10}

\bibitem{Aho:1975:ESM:360825.360855}
A.~V. Aho and M.~J. Corasick.
\newblock Efficient string matching: An aid to bibliographic search.
\newblock {\em Commun. ACM}, 18(6):333--340, June 1975.

\bibitem{1364659}
M.~Attig, S.~Dharmapurikar, and J.~Lockwood.
\newblock Implementation results of bloom filters for string matching.
\newblock In {\em Field-Programmable Custom Computing Machines, 2004. FCCM
  2004. 12th Annual IEEE Symposium on}, pages 322--323, April 2004.

\bibitem{bellekens}
X.~Bellekens, I.~Andonovic, R.~Atkinson, C.~Renfrew, and T.~Kirkham.
\newblock Investigation of {GPU}-based pattern matching.
\newblock In {\em The 14th Annual Post Graduate Symposium on the Convergence of
  Telecommunications, Networking and Broadcasting (PGNet2013)}, 2013.

\bibitem{5437720}
Y.~W. Cheng.
\newblock Fast virus signature matching based on the high performance computing
  of {GPU}.
\newblock In {\em Communication Software and Networks, 2010. ICCSN '10. Second
  International Conference on}, pages 513--515, Feb 2010.

\bibitem{cook2012cuda}
S.~Cook.
\newblock {\em {CUDA} Programming: A Developer's Guide to Parallel Computing
  with {GPU}s}.
\newblock Applications of GPU Computing Series. Elsevier Science, 2012.

\bibitem{4041182}
N.~Jacob and C.~Brodley.
\newblock Offloading ids computation to the {GPU}.
\newblock In {\em Computer Security Applications Conference, 2006. ACSAC '06.
  22nd Annual}, pages 371--380, Dec 2006.

\bibitem{kirk2012programming}
D.~Kirk and W.~Hwu.
\newblock {\em Programming Massively Parallel Processors: A Hands-on Approach}.
\newblock Elsevier Science, 2012.

\bibitem{5683320}
C.-H. Lin, S.-Y. Tsai, C.-H. Liu, S.-C. Chang, and J.-M. Shyu.
\newblock Accelerating string matching using multi-threaded algorithm on {GPU}.
\newblock In {\em Global Telecommunications Conference (GLOBECOM 2010), 2010
  IEEE}, pages 1--5, Dec 2010.

\bibitem{Matsumoto:1998:MTE:272991.272995}
M.~Matsumoto and T.~Nishimura.
\newblock Mersenne twister: A 623-dimensionally equidistributed uniform
  pseudo-random number generator.
\newblock {\em ACM Trans. Model. Comput. Simul.}, 8(1):3--30, Jan. 1998.

\bibitem{Nvidia}
Nvidia.
\newblock Cuda {C} programming guide, 2013.
\newblock http://docs.nvidia.com/cuda/.

\bibitem{Pungila:2012:HMA:2403775.2403806}
C.~Pungila and V.~Negru.
\newblock A highly-efficient memory-compression approach for {GPU}-accelerated
  virus signature matching.
\newblock In {\em Proceedings of the 15th International Conference on
  Information Security}, ISC'12, pages 354--369, Berlin, Heidelberg, 2012.
  Springer-Verlag.

\bibitem{Pungila}
C.~P. Pungila.
\newblock {\em heterogeneous pattern matching for intrusion detection systems
  and digital forensics}.
\newblock PhD thesis, West University of Timisoara, 2012.

\bibitem{Sanders:2010:CEI:1891996}
J.~Sanders and E.~Kandrot.
\newblock {\em {CUDA} by Example: An Introduction to General-Purpose {GPU}
  Programming}.
\newblock Addison-Wesley Professional, 1st edition, 2010.

\bibitem{FPGA1}
I.~Sourdis and D.~Pnevmatikatos.
\newblock Fast, large-scale string match for a 10gbps fpga-based network
  intrusion detection system.
\newblock In P.~Cheung and G.~Constantinides, editors, {\em Field Programmable
  Logic and Application}, volume 2778 of {\em Lecture Notes in Computer
  Science}, pages 880--889. Springer Berlin Heidelberg, 2003.

\bibitem{6299287}
R.~Takahashi and U.~Inoue.
\newblock Parallel text matching using {GPGPU}.
\newblock In {\em Software Engineering, Artificial Intelligence, Networking and
  Parallel Distributed Computing (SNPD), 2012 13th ACIS International
  Conference on}, pages 242--246, Aug 2012.

\bibitem{Gravity}
G.~Vasiliadis and S.~Ioannidis.
\newblock Gravity: A massively parallel antivirus engine.
\newblock In S.~Jha, R.~Sommer, and C.~Kreibich, editors, {\em Recent Advances
  in Intrusion Detection}, volume 6307 of {\em Lecture Notes in Computer
  Science}, pages 79--96. Springer Berlin Heidelberg, 2010.

\bibitem{938073}
N.~Yazdani and P.~Min.
\newblock Prefix trees: new efficient data structures for matching strings of
  different lengths.
\newblock In {\em Database Engineering and Applications, 2001 International
  Symposium on.}, pages 76--85, 2001.

\end{thebibliography}
% ACM needs 'a single self-contained file'!
%\balancecolumns
\end{document}